# Design guidelines to increase the persuasiveness of achievement goals for physical activity

Maximilian Altmeyer[a], Pascal Lessel[a], Atiq Ur Rehman Waqar[b] and Antonio Krüger[a]

[a]*German Research Center for Artificial Intelligence (DFKI), Saarland Informatics Campus, Saarbrücken, Germany*
[b]*Saarland University, Saarland Informatics Campus, Saarbrücken, Germany*

**Abstract**
Achievement goals are frequently used to support behavior change. However, they are often not specifically designed for this purpose nor account for the degree to which a user is already intending to perform the target behavior. In this paper, we investigate the perceived persuasiveness of different goal types as defined by the 3x2 Achievement Goal Model, what people like and dislike about them and the role that behavior change intentions play when aiming at increasing step counts. We created visualizations for each goal type based on a qualitative pre-study (N=18) and ensured their comprehensibility (N=18). In an online experiment (N=118), we show that there are differences in the perception of these goal types and that behavior change intentions should be considered to maximize their persuasiveness as goals evolve. Next, we derive design guidelines on when to use which type of achievement goal and what to consider when using them.

## 1. Introduction

Physical inactivity is one of the main health concerns in today's society, caused by more and more people leading sedentary lifestyles [1]. This lack of physical activity leads to a wide range of health problems, including cardiovascular diseases, obesity and numerous other chronic illnesses [2]. Consequently, encouraging people to increase their step count offers great potential to increase personal and public health [3]. Given this potential, various systems trying to encourage physical activity have been designed, implemented and studied [4]. To motivate and engage users, such systems often make use of motivational affordances such as providing visual or audio feedback [5] or gameful elements such as points or leaderboards [4]. Independent of which motivational affordances or gameful elements are used, most of them implicitly or explicitly define objectives by establishing goals that users should meet. Consequently, goals are among the most frequently used motivational affordances to induce behavior change [4, 5]. Past research has shown that the purpose for engaging in achieving such goals differs [6]. The 3x2 Achievement-Goal Model [6] explains these differences by distinguishing six types of achievement goals that are distinguished by their valence and definition of competence. However, the persuasiveness of different goal types has not been studied, as far as we know. Furthermore, research has demonstrated that the dynamic process of behavior change, i.e. a user's intention towards adopting a certain behavior, plays an important role in the perception of gameful elements for behavior change [7]. Similarly, it was shown that needs and objectives dynamically change when using fitness trackers [8], calling for dynamic adjustments of the types of goals presented to the user. We contribute to this by investigating whether a users' behavior change intention affects the perceived persuasiveness of the different types of achievement goals to inform which type of goals to offer to users as their fitness level changes. We created a set of achievement goal visualizations based on the Achievement-Goal Model [6] and an analysis of user requirements, investigate their perceived persuasiveness as well as whether it is influenced by behavior change intentions. In addition, we provide insights about what people like and dislike about them. Our contribution is two-fold: We evaluate the perceived persuasiveness and perception of the three proposed goal types as well as contribute to ongoing personalization efforts in persuasive technology research by investigating the role of behavior change intentions as a factor to consider when accounting for evolving goals and objectives. Our results show that all achievement goal types are perceived as persuasive, that there are differences in their perception and that behavior change intentions should be considered to maximize their persuasiveness. We derive five design guidelines informing the utilization of achievement goals encouraging physical activity and providing suggestions for when to use which type of goal. These design guidelines are relevant for a broad range of gameful and persuasive systems aiming at encouraging physical activity.

---

## 2. Related work

We start by presenting research in the field of persuasive technology encouraging physical activity. Next, we present research about goals and behavior change. Since we contribute to personalization research by investigating the impact of behavior change intentions, a section about personalization of persuasive systems follows.

### 2.1. Gameful and persuasive systems encouraging physical activity

UbiFit Garden [9], a system that shows a virtual garden on the wallpaper of participants' mobile phones, has been shown to lead to positive effects on their activity levels. The system establishes predefined activity goals and conveys progress towards these goals visually, through flowers and butterflies growing and appearing. Instead of using purely task-approaching goals, StepStream [10] additionally uses goals based on the performance of other users. The system shows a social stream on a website, displaying achievements when students reach their daily step goals. This adds normative feedback, allowing students to compare their own performance against others (other-approaching goal). However, the system did not lead to an increase in step counts. As reported by the authors, the intention of students to perform physical activity might have been low. Thus, using other-approach goals might have been unsuitable. In contrast, Altmeyer et al. [11] investigated the effect of showing step counts of participants on a public display in a gym, in addition to showing them in a mobile application. They found that showing each users' progress towards step goals publicly leads to a significant increase of step counts. Although the type of achievement goal differed across the aforementioned systems, it remained static within each system. However, Niess and Woźniak [8] emphasize that fitness tracker goals are not static but evolving. They introduce the "Tracker Goal Evolution Model", stating that qualitative goals (such as losing weight or doing more sports) emerge from internalized user needs. These qualitative goals can be translated into quantitative fitness goals, which can be input in a fitness tracker. However, the operationalization of how to translate qualitative goals into quantitative ones, i.e. which type of goal to use and how to cope with changing user needs and qualitative goals was not the focus of their research. We aim to contribute to this by investigating behavior change intentions as a proxy for evolving user needs and motivations.

### 2.2. Goal setting in persuasive technology and gameful systems

Ansems et al. [12] investigated the difference between using self-based goals and task-based goals regarding motivational experiences. They found that self-based goals yielded experiences related to self-improvement and enjoyment, whereas task-based goals mainly elicited experiences related to performance and competition. One of the most influential theories targeting goals and goal-setting is Locke and Latham's Goal Setting Theory [13]. Besides others, it describes effects of setting goals on subsequent goal effects and how people respond to different goal types. The authors found that goals most strongly affect user behavior when individuals are committed to them. Commitment is thereby influenced by an individual's belief that the goal can be achieved, i.e. self-efficacy. Related to this, they also highlight that when people are confronted with tasks which are complex for them, performance goals often lead to evaluative pressure and performance anxiousness. In contrast, more ambiguous goals might be more effective in this case. These findings support the main assumption of this paper, i.e. that the behavioral intention to do sports has an impact on the persuasiveness of different goal visualizations. In a literature review by Cham et al. [14], the authors provide a reference checklist for designing persuasive goals. When giving users feedback on their progress in goal completion, the paper outlines differences between performance feedback, self-comparison feedback as well as social-comparison feedback. The authors hypothesize that performance feedback might be best suited to persuade committed, motivated and skilled users, while self-comparison might be suitable for users with low self-esteem. Similarly, they suggest that social comparison might negatively affect self-esteem. The importance of adjusting fitness goals and the type of feedback was also supported by Kappen et al. [15]. They conducted an eight-week study about how gamification influences older adults in the context of physical activity and found that gamification elements and goals inherent to these elements should be adapted to the older adult's needs and preferences.

### 2.3. Personalization of gameful and persuasive systems

Although "one-size-fits-all" approaches have been shown to lead to a wide range of positive behavioral outcomes [5], research has demonstrated that such approaches often lack to account for individual motivational differences [16], resulting in suboptimal outcomes. Therefore, relevant factors moderating the perception and persuasiveness of certain strategies, gamification elements or motivational affordances have been studied. Static factors such as personality or user-/player types have been identified to have an influence on the perception of persuasive strategies. For instance, Jia et al. [17] found that personality has an influence on the perception of certain gamification elements, suggesting that adapting gami-fied systems to the personality of its users is beneficial. This is supported by findings from Orji et al. [16] who investigated the relationship between personality traits and persuasive strategies within the health domain and found similar correlations. Besides personality, user type models have been developed to tailor systems using gameful elements. For instance, the Hexad user types model by Marczewski [18] distinguishes between six different user types, has a validated questionnaire to derive a users' type [19] and has been shown to have an impact on the perception of gamification elements [7, 20].

However, the aforementioned factors are static, i.e. they usually do not change over time. Since self-efficacy, which may change dynamically for specific activities, has been shown to be one of the most important factors to complete goals [13], considering dynamic factors to personalize behavior change support systems seems important. The interplay between static and dynamic factors when personalizing gamified, persuasive systems has been investigated by Altmeyer et al. [7]. In line with this paper, they analyzed whether behavior change intentions affect which gamification elements are relevant for different Hexad user types. Indeed, they found that behavior change intentions change the perception of certain gamification elements, highlighting the importance to consider this dynamic factor. They found that the gamification elements challenge, badges, social collaboration and social competition are significantly more persuasive for people having a high intention to change their behavior. Since challenges and badges establish task-approach goals by introducing a pre-defined, static objective, we expect that task-approach goals should be more relevant for people in high stages of change, i.e. we expect a positive correlation between the stage of change and the persuasiveness of task-approach goals. Similarly, based on [7], we expect that other-approach goals should be positively correlated with the stage of change, since social strategies like competition and collaboration build on normative feedback based on the performance of other users, which is similar for other-approach goals.

## 3. Background

This section introduces relevant underlying models.

### 3.1. The 3x2 achievement-goal model

The 3x2 Achievement Goal Model [6] differentiates six types of goals by their valence (*approach, avoidance*) and their definition of competence (*task-based, self-based, other-based*):

**Task-approach**, solving a task correctly.

**Task-avoidance**, avoiding doing a task incorrectly

**Self-approach**, performing better than before

**Self-avoidance**, avoiding doing worse than before

**Other-approach**, doing better than others

**Other-avoidance**, avoiding doing worse than others

Since avoidance goals have been shown to have detrimental effects on performance, interest and task mastery [21, 22], we decided to focus on the positive valence dimension.

### 3.2. Behavior change intentions

We formalize behavior change intentions through the Transtheoretical model of behavior change by Prochaska et al. [23]. The model describes intentional behavior change as a process and bases on the assumption that behavior change involves progressing through five qualitatively different and sequential stages. These are called "stages of change" and are described in the following:

**Precontemplation:** The user has no intention to take action in the foreseeable future (6 months)

**Contemplation:** The user intends to take action within the foreseeable future (6 months)

**Preparation:** The user intends to take action in the immediate future (usually 30 days) and has taken some behavioral steps

**Action:** The user has changed their behavior for less than 6 months

**Maintenance:** The user has changed their behavior for more than 6 months

When individuals progress through these stages, their behavioral regulation becomes more self-determined [24]. We expect this to have an effect on the perceived persuasiveness of goal types.

## 4. Design process

We use the Achievement-Goal Model as a basis for the three types of goals. We started by a conducting a qualitative pre-study to elicit requirements for the realization of the goal types. After designing the three goal visualizations, we conducted another pre-study to investigate whether the visualizations are comprehensible.

### 4.1. Design requirements analysis

We conducted an online study on Prolific[1]. The study took approximately 8 minutes and participants were paid

---
[1] https://www.prolific.co/, last accessed January 30, 2021

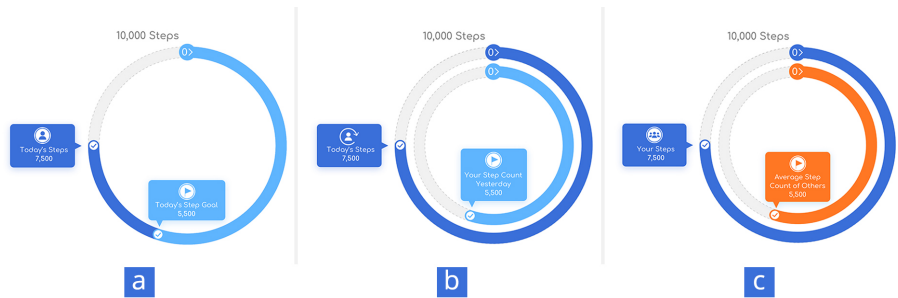

**Figure 1:** Goal visualizations for task-approach ("TAG") (a), self-approach ("SAG") (b) and other-approach goals ("OAG") (c).

GBP 1. We asked participants *"Imagine you want to increase your daily step count. How would a step goal visualization have to look like in a fitness application to motivate you reaching this goal?"*. They could enter their response in a free-text field. 18 participants took part (11 female, 7 male; age: 18-24:4, 25-31:4, 32-38:5, 39-45:2, 46-52:2, 53-59:1), of which 50% reported to do sports on a regular basis. The written answers were analyzed by conducting an inductive content analysis [25] resulting in a set of overarching themes. Based on this, we derived the following requirements for the realization of the visualizations:

**RE1: Progress**: 11 participants explicitly stated that real-time feedback is important to them.

**RE2: Visualize progress through a circular chart**: 55% of those participants that stated that real-time progress is important to them asked for visualizing step counts through graphs. Of those, 67% explicitly asked for charts using circular charts.

**RE3: Show a concrete step count**: 4 participants stated that showing a concrete step count is important to them.

**RE4: Use bright colors**: The graphs should be visualized using bright colors (mentioned by 2 participants).

### 4.2. Realization

The three types of goals to be designed are defined by the Achievement-Goal Model. In addition, the requirements **RE1** to **RE4** were considered. We decided to use circular charts to provide real-time feedback about the current step count of users (cf. **RE1, RE2**). To account for **RE3**, we decided to show a concrete step count in the visualizations. Since blue is in general a positively perceived color irrespective of demographic factors [26], we decided to use blue for visualizing progress in our designs (cf. **RE4**). To visualize the progress of other users in the other-approach goal, we decided to use the complementary of blue, orange. To realize the concept of the task-approach goal, a concrete objective is shown to the user, indicated by a respective icon and a label. To avoid confusion, the self-approach and the other-approach goal make use of a second circular chart in the inner circle of the graph to visualize one's own progress and the progress of other users respectively. To ensure the comparability of the visualizations, each one uses 5.500 steps as a goal. The final designs are shown in Figure 1.

### 4.3. Comprehensibility analysis

Next, we wanted to ensure that participants understand each type of goal. We setup an online survey on Prolific showing participants each goal visualization one by one and asking them to describe them textually. Again, the study took approximately 8 minutes and participants were paid GBP 1. The textual descriptions were analyzed by two independent raters ("RA1", "RA2"). Their task was to rate how well each goal visualization was understood on a 3-point scale (1-very poor to 3-very well). Raters were told to assign the value "0" when there is not enough information to judge whether a participant understood a certain goal visualization. The neutral choice in the comprehensibility rating was used when the main concept was understood, but specific details were either not mentioned or misunderstood.

18 participants took part (11 female, 7 male; age: 18-24:6, 25-31:6, 32-38:1, 39-45:1, 46-52:1, 53-59:2, >59:1) of which 39% reported to do sports on a regular basis. When at least one reviewer rated a description as "0", the description was not considered for the analysis. This lead to the exclusion of 17 out of 54 descriptions. To ensure that the ratings can be interpreted objectively, we calculated the inter-rater agreement and found it to be Kappa=0.85, which is considered as almost perfect [27]. Analyzing the ratings of the two independent raters, we found that the participants understood the goal visualizations very well ($M_{RA1}$ = 2.90, $Min_{RA1}$ = 2; $M_{RA2}$ = 2.96, $Min_{RA2}$ = 2). Based on this, the three goal visualizations could be used to investigate their perception and perceived persuasiveness in the main study.

## 5. Evaluation

We investigate the persuasiveness of the goal types, reasons and the role of behavior change intentions.

### 5.1. Hypotheses

We expect to find evidence for the following hypotheses:

**H1**: The perceived persuasiveness and user preferences differ between the task-, self, and other-approach goals.

**H2**: Task-approach goals are perceived as more persuasive among people in high stages of change.

**H3**: Other-approach goals are perceived as more persuasive among people in high stages of change.

**H4**: Self-approach goals are perceived as more persuasive among people in low stages of change.

We expect that the visualizations are perceived differently (**H1**). This is supported by findings from [12], showing that self-based goals were more focused on improvement than on performance. **H2** is motivated by task-based goals affording a certain perceived competence to reach them and were shown to elicit responses related to performance [12]. **H2** is supported further by findings by Cham et al. [14] and Locke and Latham [13], showing that performance is associated with motivation and skill and that self-efficacy plays a major role in goal attainment. Similarly, **H3** is motivated by the importance of self-efficacy for the relevance of goals. Since comparing to other users might establish normative standards which seem to be out of reach for users in low stages of change, we expect that other-approach goals should be more relevant for users in high stages. This is supported by findings from Altmeyer et al. [7], showing that social gamification elements are more suitable for users in high stages of change. In contrast, establishing goals based on one's own performance should lead to reachable goals, which might be more suitable for users with lower self-esteem [14] in low stages of change (**H4**).

### 5.2. Method and procedure

We conducted an online experiment on Prolific. It took approximately 12 minutes to complete and was approved by our Ethical Review Board [2] (#20-02-4). Participants were paid GBP 1.50. After asking for demographic data, the stage of change was determined using a validated scale for the physical activity context [28]. Next, participants were shown each of the three goal visualizations individually in a random order and were asked to fill out the validated perceived persuasiveness scale by Thomas, Masthoff and Oren [29]. The scale consists of three factors (effectiveness, quality, capability) measured on 7-point scales. In addition to that, we also asked participants to describe what they like and dislike about the presented visualizations in a mandatory free-text field. The textual responses were qualitatively analyzed in order to understand *why* participants perceived the visualizations as persuasive or not. We analyzed the responses systematically by conducting an inductive content analysis [25] to identify patterns of meaning (themes). After being shown each of the three goal visualizations individually, participants were shown all visualizations at once (next to each other) and asked to select which of them they personally like the most to assess their overall preference. A Shapiro-Wilk test revealed that the persuasiveness scale items were not normally distributed, which is why we used non-parametric tests for our analysis. For correlation analysis, Kendall's $\tau$ was used, as it is well-suited for non-parametric data [30]. Kendall's $\tau$ is usually lower than Pearson's r for the same effect sizes. Therefore, we transformed interpretation thresholds for Pearson's r to Kendall's $\tau$, according to Kendall's formula [31] (small: $\tau = 0.2$; medium: $\tau = 0.3$; large: $\tau = 0.5$).

### 5.3. Results

|  | | **TAG** Task-Approach | **SAG** Self-Approach | **OAG** Other-Approach |
|---|---|---|---|---|
| | Preference | M=0.25 SD=0.44 Mdn=0.00 | M=0.44 SD=0.50 Mdn=0.00 | M=0.31 SD=0.46 Mdn=0.00 |
| Perceived Persuasiveness Scale | Persuasiveness | M=5.20 SD=0.94 Mdn=5.33 | M=5.29 SD=0.94 Mdn=5.44 | M=5.15 SD=0.98 Mdn=5.33 |
| | Effectiveness | M=4.83 SD=1.37 Mdn=5.00 | M=4.98 SD=1.35 Mdn=5.33 | M=4.75 SD=1.45 Mdn=5.00 |
| | Quality | M=5.26 SD=1.00 Mdn=5.33 | M=5.31 SD=0.99 Mdn=5.33 | M=5.05 SD=1.07 Mdn=5.00 |
| | Capability | M=5.50 SD=1.07 Mdn=6.00 | M=5.59 SD=0.98 Mdn=6.00 | M=5.64 SD=1.08 Mdn=6.00 |

**Table 1**
Descriptive data of dependent variables for each goal visualization. TAG=Task-Approach Goal, SAG=Self-Approach Goal, OAG=Other-Approach Goal

We excluded participants who answered one of three test questions incorrectly, leading to a final answer set of 118 responses (64 female, 54 male; age: 18-24: 24, 25-31: 38, 32-38: 18, 39-45: 19, 46-52: 9, 53-59: 6, >59: 4). 12 participants were in the precontemplation, 18 in the contemplation, 28 in the preparation, 20 in the action and 40 in the maintenance stage of change. 49 participants were not doing any kind of sports whereas 69 did.

---
[2]https://erb.cs.uni-saarland.de/, last accessed January 30, 2021

### 5.3.1. Differences between goal visualizations

The mean and median scores for all dependent variables can be found in Table 1. All goal visualizations were perceived as persuasive, as revealed by one-sample Wilcoxon signed rank tests against the neutral choice of four on the 7-point scale. The persuasiveness score is significantly higher than four for all goal types (each p<0.001). This leads to result **R1: All goal visualizations are perceived as persuasive**. Next, we analyzed whether there are differences between the goal visualizations. We calculated a Friedman's ANOVA for all dependent variables shown in Table 1 and used the Durbin-Conover method for post-hoc analysis. The Benjamini-Hochberg false discovery rate [32] was used to adjust significance values for multiple comparisons. We found a significant effect for the user preferences ("Preference") ($\chi^2(2)=6.58$, p<0.05) and for the "Quality" factor of the perceived persuasiveness scale ($\chi^2(2)=15.20$, p<0.01). For all other variables, no effects were found. The post-hoc analysis revealed **R2: Participants prefer self-approach goals over task-approach goals** (p=0.039). Regarding the "Quality" factor of the perceived persuasiveness scale, relating to the trustworthiness of the goal visualizations, pairwise comparisons revealed that **R3: Participants considered self-approach goals as more trustworthy than other-approach goals** (p=0.003) as well as **R4: Participants considered task-approach goals as more trustworthy than other-approach goals** (p=0.003). No further significant differences were found.

### 5.3.2. Effect of the stage of change

To analyze the effect of the stage of change on the perceived persuasiveness of the three goal visualizations, we analyzed whether the a-priori formulated relationships (cf. **H2–H4**) exist by calculating one-tailed bivariate correlations between the stage of change and the items measuring perceived persuasiveness. We found that the overall persuasiveness of both the task-approach ($\tau$=.18, p<.01) and other-approach goals ($\tau$=.16, p<.05) are positively correlated with the stage of change. More specifically, we found that the "Effectiveness" ($\tau$=.16, p<.05) and "Capability" ($\tau$=.16, p<.05) factor of task-approach goals are positively correlated with the stage of change. In sum, we formulate **R5: Task-approach goals are perceived as more persuasive with increasing stages of change, mostly because of higher "Effectiveness" and "Capability" scores**. For other-approach goals, we found a positive correlation for the "Quality" factor ($\tau$=.17, p<.01), leading to **R6: Other-approach goals are perceived as more persuasive with increasing stages of change, mostly because of a higher "Quality"**. For self-approach goals, no correlations were found.

### 5.3.3. Qualitative analysis

To better understand the underlying reasons for the quantitatively found effects, we analyzed the textual responses by participants qualitatively. For each participant and visualization, one textual response was recorded, resulting in 354 responses that have been analyzed. First, we analyzed whether participants perceived the goal visualization negatively, neutral or positively by assigning values from 1–3 to each textual summary (1=negative, 2=neutral, 3=positive). On average, participants were rather neutral about task-approach (M=2.08, SD=0.49), self-approach (M=2.14, SD=0.60) and other-approach goals (M=1.92, SD=0.69). A Friedman ANOVA revealed that there is a significant difference between these values (p=0.043). Pairwise comparisons using the Durbin-Conover method and the Benjamini-Hochberg false discovery rate [32] revealed that self-approach goals were coded to be perceived more positively than other-approach goals, which is similar to **R3**. No effects were found between task-approach goals and other approach goals.

When analyzing what participants like and dislike about the goal types, several themes emerged that might explain the results that have been found based on the quantitative analysis. Themes are written in bold italics. First, the fact that participants in general preferred self-approach goals over task-approach goals (**R2**) seems to be related to participants considering self-approach goals as ***meaningful***. This theme was found consistently across the free text answers about aspects that participants liked about self-approach goals. Self-approach goals are considered as "personally relevant" (**P115**), and considered to "give you a reason to push yourself against your own goals [...]" (**P115**). In contrast, task-approach goals were considered as ***arbitrary*** or ***meaningless***. Participants reported that "it lets you set an arbitrary goal that may not mean as much" (**P105**). A main reason for why participants considered self-approach goals as more meaningful seems to be related to ***self-improvement***. Participants said that "I like that it shows I am improving so I'd feel good about that" (**P73**) or that "seeing this makes me want to improve upon my previous record" (**P104**). In addition, participants consider self-approach goals as more ***healthy***, i.e. they liked that the self-approach visualization establishes moderate, reachable goals. **P25** notes that "it can work slowly towards achieving the goal". This is supported by **P47**, stating that "it can make me more competitive with myself in a fun and healthy way".

Regarding **R3** and **R4**, trust seems to play a major role. This is supported by the thematic analysis, revealing that the major drawback of other-approach goals is missing ***trust***. We found that participants were afraid that other users might ***cheat*** to increase their step counts or that ***technology*** is not capable to reliably measure the steps taken. **P67** notes that other people "may not

*be doing the right thing anyway"* and **P68** states that *"I would doubt the accuracy of the data [...] or want more information about where it comes from"*. In addition, our analysis revealed that **comparability** is a major concern. Participants frequently stated that they do not have enough information to judge whether others are comparable in terms of their **fitness level**, their **demographics** or their **circumstances**. A statement by **P77** summarizes this: *"Circumstances are different for everyone. The people who walk more could have more time on their hands, could be walking a lot in their work so it wouldn't influence me to exercise more"*. In addition, participants were concerned about **over-training** when using the other-approach goal. They noted that seeing other users' step counts might lead to peer-pressure which may result in people doing more than is good for them. **P79** states that *"My targets are based on my health needs and not on what others are doing"* and emphasizes that the other-based goal *"would encourage me to do more than my limbs may be ready for by tapping into my competitive spirit"*. However, in line with **R1**, participants also reported positive aspects about task-approach and other-approach goals. They like that task-approach goals are **objective**, **simplistic** and consider them as **reliable**. **P99** states that *"I like how accurate it is"* and **P78** supports this by stating that *"It is motivating and something concrete to base exercise on"*. Also, participants like that task-approach goals are not related to one's own performance and thus are perceived as rather **non-binding**: **P81** states that *"It encourages the reaching of goals without worrying about shortfalls prior to the current day"* which is supported by **P71** stating that *"I like the simplistic approach and that there isn't a comparison to anything, it's just your step count and whether you have beaten your daily goal"*. For other-approach goals, participants liked that it may push **self-efficacy** when one's own performance is better than those of others: *"it leads me to believe that I am achieving above average results which makes me feel good about myself"* (**P64**). Participants also reported that **competitiveness** is a strong motivator for them: *"I like that it keeps people competitive"* (**P87**).

## 6. Discussion

Our results show that in general, all three goal types are perceived as persuasive (**R1**). This might be related to the fact that all visualizations establish goals, which has been shown to affect action [13]. However, differences between the three goal visualizations were found. First, we found that participants preferred self-approach goals over task-approach goals (**R2**). Based on the qualitative analysis, it seems that participants appreciated that self-approach goals support self-improvement and thus considered these goals as more meaningful. This is in line with findings from Ansems et al. [12] who compared task-based against self-based goals in a dance game and found that participants responded more in terms of self-improvement in the self-based condition. Second, we found that both self-approach and task-approach goals were perceived as more trustworthy than other-approach goals, as revealed by significantly higher scores on the "Quality" factor of the perceived persuasiveness scale (**R3, R4**). When analyzing reasons for what participants did not like about other-approach goals qualitatively, we found supporting evidence for this effect, since trust emerged as a main theme. When further unfolding this, we learned that participants did not trust the data of other users mainly because they expected them to cheat and because they are concerned about measuring errors of step counters. These findings are in line with results by Niess and Woźniak [8] who found that building trust in the goal and in the fitness tracker is important for the goal to be meaningful. Thus, taking **R2–R4** together, **H1: The perceived persuasiveness and user preferences differ between the task-, self, and other-approach goals** is partially supported.

We furthermore learned that the stage of change is positively correlated with the overall perceived persuasiveness of task-approach goals. The "Effectiveness" and the "Capability" factors are positively correlated with the perceived persuasiveness of task-approach goals (**R5**). Given that objectiveness emerged as a main theme when analyzing what participants liked about task-approach goals, it seems that participants appreciated to have a clear goal which allows attaining task-based competence. This seems to be a reasonable explanation for why there is a positive relationship between the stage of change and the perceived persuasiveness of task-approach goals when considering findings by Altmeyer et al. [7]. They show that gamification elements such as challenges and badges, establishing clear goals allowing to evaluate how well or badly a task was solved, were perceived as significantly more motivating by participants in high stages of change. As such, we consider our results as supporting evidence for **H2: Task-approach goals are perceived as more persuasive among people in high stages of change**. We found that the perceived persuasiveness of other-approach goals is positively correlated with the stage of change of participants. Besides finding a positive correlation between the overall perceived persuasiveness and the stage of change, we also found a positive correlation between the "Quality" factor of the perceived persuasiveness scale and the stage of change (**R6**). Thus, it seems like the perceived trustworthiness, which is measured by the "Quality" factor [29], is the deciding cause for this positive correlation. Again, this is supported by the qualitative analysis, revealing that missing trust (mostly because of expecting other users to cheat and a low perceived accuracy of technology) is a main theme that emerged. Given that low ability to perform a task has been shown to be a key factor for cheating behavior [33] the positive correlation to the stage of change seems reasonable. Additionally, the positive correlation between the stage of change and the perceived persuasiveness of the other-approach goal is in line with [7], showing that leaderboards are perceived as significantly more persuasive by users in high stages of change. These results support **H3: Other-approach goals are perceived as more persuasive among people in high stages of change**. When analyzing a potentially negative correlation between the stage of change and the perceived persuasiveness of self-approach goals, we could not find significant effects. It seems like self-approach goals are perceived positively across all stages of change. This is in line with findings from the thematic analysis, since no themes emerged in this regard. The fact that self-approach goals adapt to the personal performance of participants, which in turn encourages intrinsic motives such as self-improvement, seems to stimulate both participants in high and low stages of change. Again, this seems to relate to findings by Altmeyer et al. [7], who found that both people in low and high stages of change perceived personalization, i.e. a system adapting the step goal to individuals, positively. Thus, based on our results, we did not find evidence for **H4: Self-approach goals are perceived as more persuasive among people in high stages of change**.

### 6.1. Design guidelines

Based on both the quantitative and qualitative results, we establish the following set of design guidelines:

**Use self-approach goals when having no information about a user's stage of change**

Our results show that self-approach goals are perceived as persuasive and are preferred over task-based and other-based goals, independent of the stage of change of users (cf. **R1, R2, R3**). Therefore, we generally recommend to use self-approach goals in systems encouraging physical activity in order to support self-improvement, which was shown to be perceived as meaningful by participants. Also, this type of goal was considered as healthy, since it is based on one's own performance.

**Use task-approach or other-approach goals for users in high stages of change**

The findings (**R5**) show that task-approach goals are more relevant for people in higher stages of change. The fact that task-based goals establish a clear objective and allow to easily evaluate whether or not it has been met seems to be the deciding factor for the positive perception. Instead of focusing on self-improvement, task-based goals focus on mastery [12], which has been found to be more relevant for users that internalized their behavior [7].

Regarding other-approach goals, our results show that they should be more relevant for users in high stages of change (**R6**) and that users like other-approach goals mainly due to the inherent competitiveness of this type of goals. That competition is more relevant for users in high stages of change has also been shown in previous research [7], supporting our findings.

**For self-approach goals, support self-improvement**

Through the thematic analysis, we found that self-improvement is a strong motivator when using self-approach goals as it is perceived as meaningful. Therefore, we recommend to focus on supporting self-improvement when using self-approach goals. This could be achieved by highlighting personal growth, e.g. by visualizing trends of physical activity over time or by introducing metrics making self-improvements more graspable such as showing the relative improvement over a certain timespan.

**For task-approach goals, avoid arbitrariness**

Participants appreciated that task-based goals introduce a clear, reliable target which can be objectively measured. However, there is as risk that this might seem arbitrary and thus meaningless to users. Therefore, we suggest to make these goals more meaningful, by adding personal relevance through comparisons to the real world that make an arbitrary number more graspable. This could be achieved by comparing the step goal to a distance that people might be able to relate to (such as the distance between two cities that are known to users).

**For other-approach goals, focus on transparency, comparability and avoiding over-training**

Missing trust emerged as a main theme. This is in line with our quantitative results (**R3, R4**). Therefore, we recommend to communicate and explain transparently how the data of others has been measured and aggregated. Moreover, since comparability (in terms of demographics or fitness level) has been raised as another major concern, we recommend to provide information about the sample that the individual is compared to or even select a subset of other users which is comparable to the individual in terms of fitness level and demographics. Moreover, as revealed by the qualitative analysis, measures to prevent over-training should be incorporated.

### 6.2. Limitations

We used static visualizations to assess the perceived persuasiveness of each goal type; we did not implement them. Although this has several advantages such as reaching a higher number of participants and abstracting from specific implementation choices, which could bias the results in a way that is hard to control [34], validating our findings using real implementations is an important next step. Also, the fact that we decided to show visualizations in which participants had already reached their goal, might affect the perceived persuasiveness of the goal visualizations. Although we tried to inform the realization of the three goal types by a pre-study and showed that the realizations were successfully illustrating the intended goal types by a second pre-study, it should be noted that there might be other realizations of the types of goals leading to different results. Lastly, it should be considered that we used Prolific, a platform for paid online studies, which might have affected our results.

## 7. Conclusion and future work

We designed three different goal visualizations based on the 3x2 Achievement-Goal model to encourage physical activity. We first elicited requirements that should be considered in a pre-study (N=18), realized the three visualizations and made sure that they are understood by participants as well as visualize the intended concepts in a second pre-study (N=18). Next, we investigated the perceived persuasiveness of each goal visualization, the effect of behavior change intentions and which aspects participants like or dislike.

We found that all goal types are perceived as persuasive. Self-based goals were preferred over task-based goals by participants and were perceived as more trustworthy than other-approach goals. Moreover, task-based goals were considered as more trustworthy than other-approach goals, too. Furthermore, we found that behavior change intentions have an effect on the perceived persuasiveness of both the task-approach as well as the other-approach goal types.

Our qualitative analysis revealed that participants like that task-approach goals are reliable, simple and objective whereas they dislike their arbitrariness. They liked that self-approach goals support self-improvement, which was considered as meaningful. For other-approach goals, participants appreciated their competitiveness which may boost self efficacy whereas they were concerned about the trustworthiness of other people's data and the missing comparability in terms of fitness level or demographics. Based on these results, we derived design guidelines that support the development and design of persuasive systems aiming at encouraging physical activity. In future work, concrete implementations of the three goal types increasing the step counts of users should be used to investigate whether our results can be replicated when measuring actual user behavior. Also, other contexts should be investigated as well as whether our findings can be transferred to avoidance goals.